\documentclass[pra,showpacs,superscriptaddress,preprintnumbers,twocolumn,amsmath,amssymb,floatfix]{revtex4}

\usepackage{graphicx}
\usepackage{bm}
\usepackage{epsfig}
\usepackage{dcolumn}
\usepackage{pstricks,,pst-node,pst-tree}

\newcommand{\expv}[1]{\ensuremath{\langle #1  \rangle}}

\begin{document}

\preprint{submitted to Phys.\ Rev.\ A}

\title{Self-interaction in Green's-function theory of the hydrogen atom}

\author{W. Nelson}
\altaffiliation[Present address: ]{Department of Physics, King's College 
London, Strand,
London WC2R 2LS, United Kingdom}
\affiliation{Department of Physics, University of York, Heslington, York
             YO10 5DD, United Kingdom}
\author{P. Bokes}%
\affiliation{Department of Physics, Faculty of Electrical Engineering and
             Information Technology, Slovak University of Technology,
             Ilkovi\v{c}ova 3, 841 04 Bratislava, Slovak Republic}
\affiliation{European Theoretical Spectroscopy Facility (ETSF)}
\author{Patrick Rinke}
\affiliation{European Theoretical Spectroscopy Facility (ETSF)}
\affiliation{Fritz-Haber-Institut der Max-Planck-Gesellschaft,
             Faradayweg 4--6, 14195 Berlin, Germany}
\author{R. W. Godby}
\email{rwg3@york.ac.uk}
\affiliation{Department of Physics, University of York, Heslington, York
             YO10 5DD, United Kingdom}
\affiliation{European Theoretical Spectroscopy Facility (ETSF)}

\date{\today}

\begin{abstract}

Atomic hydrogen provides a unique test case for computational electronic 
structure
methods, since its electronic excitation energies are known analytically.
With only one electron, hydrogen contains no electronic correlation and is
therefore particularly susceptible to spurious self-interaction errors
introduced by certain computational methods.
In this paper we focus on many-body perturbation-theory (MBPT)
in Hedin's $GW$ approximation. While the Hartree-Fock and
the exact MBPT self-energy are free of self-interaction, the correlation 
part
of the $GW$ self-energy does not have this property.
Here we use atomic hydrogen as a benchmark system for $GW$ and show that
the self-interaction part of the $GW$ self-energy, while non-zero, is small. 
The effect of calculating the $GW$ self-energy from exact 
wavefunctions and eigenvalues, as distinct from those from the local-density 
approximation, is also illuminating.
\end{abstract}

\maketitle

\section{INTRODUCTION}
\label{section.intro}

{\it Ab initio} many-body quantum mechanical calculations are crucially
important to our understanding of the behavior of atomic, molecular
and condensed matter systems.
It is well known that
prediction of the behavior of these systems requires the description of 
electronic
correlation. Whilst density-functional theory (DFT) in the local-density
approximation (LDA) does this with startling success in many cases, it does 
so at the expense of a non-physical electron self-interaction.
For delocalized electron systems this
self-interaction becomes negligible, but in
atomic or strongly localized electronic systems
it plays an important role.
If one is interested in the calculation of quasiparticle excitation spectra,
many-body perturbation-theory (MBPT) is formally a correct way to proceed.
For solids, MBPT in Hedin's $GW$ approximation \cite{Hedin:1965} has
become the method of choice, but it is also increasingly being applied to
molecular systems and clusters.
The $GW$ self-energy can be decomposed into correlation and exchange parts, 
where the latter is the same as the Fock operator encountered in 
Hartree-Fock theory and thus self-interaction free.
While the exact self-energy must also be free of self-interaction, the
correlation part of the $GW$ self-energy does not have this property.
To investigate the influence of self-interaction in the $GW$
approach the hydrogen atom provides an ideal case, because the
exact solution is known analytically.

Hydrogen in its solid phase has previously been studied within the $GW$ 
approximation by
Li {\it et al.}~\cite{Li/etal:2002}, who analyzed the transition between the
high-pressure solid phase and the low density, atomic-like limit.
For individual atoms, $GW$ electron removal and addition energies (we use 
the term ``quasiparticle'' energies by analogy with the solid-state 
situation) have been investigated by
Shirley and Martin \cite{Shirley/Martin:1992},
Dahlen {\it et al.} \cite{Dahlen/Barth:2004,Dahlen/Leeuwen:2005},
Stan {\it et al.} \cite{Stan/Dahlen/Leeuwen:2006}
and Delaney {\it et al.} 
\cite{Delaney/Garcia-Gonzalez/Rubio/Rinke/Godby:2004},
although hydrogen was not considered.
These studies have shown that $GW$, in general, gives quasiparticle 
properties which are
much improved over DFT and Hartree-Fock methods, even for atoms.

In this paper we use the hydrogen atom as a benchmark system to quantify the 
self-interaction error in the $GW$ approach.
Since the self-energy diagrams {\it beyond} $GW$, known as the vertex 
correction, must by definition correct this self-interaction error, our 
findings are relevant for research into vertex functions for the 
many-electron problem.

Attention has recently focused on the prospects for improving the usual 
non-self-consistent
$GW$ calculations by choosing an initial Green's function, $G_0$, that is 
physically more
reasonable than the LDA (e.g. \cite{Li/etal:2002,Rinke/etal:2005,Li/etal:2005}).  
We explore this here by determining
the sensitivity of the self-interaction error to the use of the exact 
hydrogenic
orbitals and energies in place of those from the local-density approximation 
(LDA).  We also
assess the error introduced into $GW$ calculations by employing first-order 
perturbation theory in solving the quasiparticle equation (as opposed to the 
full numerical solution), and we analyze
the quasiparticle wavefunctions that emerge from a full solution.

\section{Hartree-Fock vs. DFT-LDA}
\label{section.hartree-fockvslda}

In many-body perturbation theory the quasiparticle excitation
energies $\epsilon_{i\sigma}$ and wavefunctions $\psi_{i\sigma}$
are the solutions of the quasiparticle equation
\begin{equation}
\label{qpe:full}
  H_0({\bf r})\:\psi_{i\sigma}({\bf r}) + \sum_{\sigma'}
  \int d{\bf r'} M_{\sigma\sigma'}({\bf r},{\bf r'};\epsilon_{i\sigma}^{\rm 
qp})
  \psi_{i\sigma'}({\bf r'})
   = \epsilon_{i\sigma}^{\rm qp} \psi_{i\sigma}({\bf r})
\end{equation}
where, in Hartree atomic units, $H_0({\bf 
r})=-\frac{1}{2}\nabla^2+v_{ext}({\bf r})$ and $v_{ext}({\bf r})$
is the external potential.
It is customary to divide the mass operator $M$ into the local Hartree
potential ($v_H$) and the non-local self-energy ($\Sigma$)
\begin{equation}
\label{eq:M}
  M_{\sigma\sigma'}({\bf r},{\bf r'};\epsilon)=
      v_H({\bf r})\delta({\bf r}-{\bf r'})\delta_{\sigma\sigma'}
      +\Sigma_{\sigma\sigma'}({\bf r},{\bf r'};\epsilon).
\end{equation}
Omitting correlation contributions from $\Sigma$ yields the exact-exchange 
or
Hartree-Fock case, where the self-energy takes the form
\begin{equation}
\label{eq:HF}
  \Sigma_{\sigma\sigma'}^x({\bf r},{\bf r'}) = -\sum_{i}^{occ}
                      \frac{\psi_{i\sigma}({\bf r})
		            \psi_{i\sigma}^*({\bf r'})}{|{\bf r}-{\bf r'}|} 
\delta_{\sigma\sigma'}
\end{equation}
and the Hartree potential is given by
\begin{equation}
\label{eq:vH}
  v_{H}({\bf r}) =
  \sum_{i,\sigma}^{occ}
  \int d{\bf r'}\frac{\psi_{i\sigma}({\bf r'})
                      \psi_{i\sigma}^*({\bf r'})}{|{\bf r}-{\bf r'}|}.
\end{equation}
Since the sum runs over all occupied states the Hartree potential contains 
an artificial
interaction of electrons with themselves. For the hydrogen atom this 
so-called
self-interaction of the electron is the {\it only} content of the Hartree 
potential and may be calculated
analytically using the exact ground-state wavefunction
\begin{equation}
\label{eq:H_wfc}
\psi_{1s}^{\uparrow}(r)=\frac{1}{\sqrt{\pi}} e^{-r}
\end{equation}
as
\begin{equation}
v_{H}(r)=\frac{1}{r}\left\{ {1 - \left( {1 + r} \right)e^{ - 2r} } \right\} 
.
\end{equation}
The self-interaction is a positive, decreasing function and hence tends to 
delocalize the wavefunction (i.e.
incorrectly pushes its weight away from the nucleus).

In Hartree-Fock the self-interaction terms introduced in the Hartree 
potential are exactly canceled by the Fock operator $\Sigma_x$. This makes 
Hartree-Fock exact for one-electron systems such as the hydrogen atom.
However, the lack of correlation renders Hartree-Fock unsuitable for many 
polyatomic systems of interest.

Of more practical use is Kohn-Sham \cite{Kohn/Sham:1965} density-functional
theory, which by virtue of the Hohenberg-Kohn theorem 
\cite{Hohenberg/Kohn:1964} establishes an exact and
universal functional relationship between the ground-state density 
$n({\bf r})$ and the total energy $E$ of a system. Mapping the system of 
interacting electrons onto a fictitious system of non-interacting electrons, 
that reproduces the exact density, yields the Kohn-Sham equations:
\begin{equation}
\label{eq:KS}
  \left[H_0({\bf r})+v_{H}({\bf r})+v_{xc}({\bf r}) \right]
   \phi_{i\sigma}({\bf r})
    =\epsilon_{i\sigma}^{\rm KS} \phi_{i\sigma}({\bf r}).
\end{equation}
All electron-electron interactions beyond the Hartree mean field are 
encompassed by
the exchange-correlation potential $v_{xc}$, which is formally given as the 
functional
derivative of the exchange-correlation energy $E_{xc}$:
\begin{equation}
\label{eq:vxc}
  v_{xc}({\bf r})=\frac{\delta E_{xc}[n]}{\delta n({\bf r})} \quad .
\end{equation}
In analogy to the quasiparticle equation (\ref{qpe:full}) the Kohn-Sham
eigenvalues $\epsilon_{i\sigma}^{\rm KS}$ are often interpreted as 
excitation energies,
although this is not formally justified.

One of the most common approximations for $ E_{xc}$ is the local-density 
approximation
(LDA) \cite{Kohn/Sham:1965}, in which the
many-body exchange and correlation contributions to the total energy
($E_{xc}$) are included by comparison with the homogeneous electron gas 
(HEG):
\begin{equation}
  E_{xc}^{\rm LDA}[n]=\int \! d\textbf{r} \: n(\textbf{r})
                      \epsilon_{xc}^{\rm HEG} [n(\textbf{r})]\quad.
\end{equation}
Here \cite{SLDA} we follow the parameterization of Perdew and Zunger
\cite{Perdew/Zunger:1981} for the exchange-correlation
energy density $\epsilon_{xc}^{\rm HEG}[n({\bf r})]$ of the
homogeneous electron gas  based on the data of Ceperley and Alder
\cite{Ceperley/Alder:1980}.

The LDA has been remarkably successful at accounting for
correlation even in systems that
are highly inhomogeneous. However, it is well known that
the introduction of correlation in the LDA comes at the expense of the exact
treatment of the self-interaction. Because the exchange functional is taken
from the homogeneous electron gas it no-longer
cancels the spurious self-interaction present in the Hartree term. In most
systems this is a minor
effect and is more than compensated by the improved treatment of the 
electron
correlation.  The LDA can be improved by explicitly removing the
self-interaction \cite{Perdew/Zunger:1981}, however this becomes 
increasingly
difficult as the system's complexity is increased.

\section{The \textit{GW} approximation}
\label{section.gwapprox}

In Hedin's $GW$ approximation \cite{Hedin:1965} the self-energy in
Eq.~\ref{eq:M} is given by
\begin{equation}
\label{Eq:S=GW}
  \Sigma_{\sigma\sigma}({\bf r},{\bf r'};\epsilon)=
   \frac{i}{2\pi}\int_{-\infty}^\infty \!\!\!\!\!\!
   d\epsilon' e^{i\epsilon'\delta}
    G_{\sigma\sigma}({\bf r},{\bf r'};\epsilon+\epsilon')
    W({\bf r},{\bf r'};\epsilon)
\end{equation}
where $\delta$ is an infinitesimal positive time.
At the level of $GW$, spin flips are not accounted for
\cite{Li/etal:2002}, and the input Green's
function is diagonal in its spin representation
$G_{\sigma\sigma}=G_{\sigma\sigma'}\delta_{\sigma\sigma'}$

In common with usual $GW$ calculations, a Kohn-Sham Green's function $G^0$ 
is used for $G$, given by \cite{delta}
\begin{equation}
\label{eq:eigenexpansion}
  G_{\sigma\sigma}^0(\textbf{r},\textbf{r}',\epsilon)=\sum _{i}
  \frac{\phi_{i\sigma}(\textbf{r})
   \phi^{*}_{i \sigma}(\textbf{r}')}
  {\epsilon-\epsilon_{i\sigma}^{\rm KS} \mp i\delta},
\end{equation}
and makes the non-interacting polarizability
\begin{equation}
\label{Eq:chi_0}
  \chi_{\sigma\sigma}^0({\bf r},{\bf r}',\epsilon)=
                -\frac{i}{\pi}\int_{-\infty}^\infty \!\!\!\!\!\!\! 
d\epsilon'
                G_{\sigma\sigma}^0({\bf r},{\bf r}';\epsilon'-\epsilon)
		G_{\sigma\sigma}^0({\bf r}',{\bf r};\epsilon')
\end{equation}
and the dielectric function spin-dependent (though spin-diagonal).
The inverse dielectric function in the random-phase
approximation
\begin{equation}
  \varepsilon^{-1}(\textbf{r},\textbf{r}',\epsilon)=\left[
               \delta({\bf r}-{\bf r}')-
		\int d{\bf r}'' v({\bf r},{\bf r}'')\sum_\sigma
	       \chi_{\sigma\sigma}^0({\bf r}'',{\bf r}',\epsilon)\right]^{-1}
\end{equation}
and thus the screened Coulomb interaction
\begin{equation}
  W_0(\textbf{r},\textbf{r}',\epsilon)=\int d{\bf r}''
               \varepsilon^{-1}({\bf r},{\bf r}'',\epsilon) v({\bf r}'',{\bf
	       r}').
\end{equation}
then emerge as spin-independent quantities, giving rise to the simple spin 
dependence in the
$GW$ self-energy (Eq.~\ref{Eq:S=GW}).

For numerical convenience and physical insight we separate the $GW$
self-energy (\ref{Eq:S=GW}) according to
\begin{equation}
  \Sigma_{\sigma\sigma}=-i[G_{\sigma\sigma}v+G_{\sigma\sigma}(W-v)]
                       =\Sigma_{\sigma\sigma}^x+\Sigma_{\sigma\sigma}^c 
\quad.
\end{equation}
The first term ($\Sigma_{\sigma\sigma}^x$) corresponds to the Fock operator 
in Eq.~\ref{eq:HF} and will exactly cancel the self-interaction introduced by 
the Hartree potential. It is therefore immediately clear that any deviation 
from the exact result for
hydrogen can only come from the correlation part of the self-energy
($\Sigma_{\sigma\sigma}^c$).

The incorrect self-interaction affects the electron removal energies (here, 
the ionization potential).  For electron addition energies such as the 
electron affinity, the entire Hartree potential has a physically reasonable 
interpretation, since it acts on the wavefunction of the originally unoccupied
state which has not contributed to the electron density.

\section{Computational Approach}
\label{sec:CA}

We solve the quasiparticle equation (\ref{qpe:full}) with the $G_0W_0$ 
self-energy
(\ref{Eq:S=GW}) for the quasiparticle energies and wavefunctions by
fully diagonalizing the quasiparticle Hamiltonian in the basis of the
single particle orbitals of the non-interacting system.
Since the ground state of the hydrogen atom (Eq. \ref{eq:H_wfc}) is 
spherically
symmetric, it is sufficient to describe all non-local operators in the $GW$
formalism  by two radial and one spin coordinates, $r,r',\sigma$
and one angular coordinate, $\theta$, that denotes the angle
between
the vectors ${\bf r}$ and ${\bf r'}$. The self-energy (Eq. \ref{Eq:S=GW}) 
then
assumes the much simpler form
\begin{equation}
  \label{eq:sigma_legexp}
  \Sigma_{\sigma\sigma}(r,r',\theta;\omega)=\sum_{l=0}^{\infty} \left[
                              \Sigma_{l\sigma}(r,r';\omega)\right]
                               P_l(\cos\theta) \delta_{\sigma,\sigma}
\end{equation}
where $P_l(\cos\theta)$ is a Legendre polynomial of order $l$.

The Legendre expansion coefficients of the self-energy are calculated 
directly,
thereby surpassing the need for a numerical treatment of the angular
dependence.
We use a real-space and imaginary time representation 
\cite{Rojas/Godby/Needs:1995}
to calculate the self-energy from the non-interacting Green's function 
$G_0$.
The expression for the self-energy on the real frequency axis is obtained by
analytic continuation \cite{Rojas/Godby/Needs:1995}. The current 
implementation
has been successfully applied to jellium clusters 
\cite{ClusterImStates:2004} and
light atoms \cite{Delaney/Garcia-Gonzalez/Rubio/Rinke/Godby:2004}.

Our code allows us to solve the quasiparticle equation (\ref{qpe:full}) 
for the $GW$ self-energy with no further approximation.  
However, in order to separate the contribution 
that arises from the correlation
part of the self-energy from that of the exchange part and the Hartree and
exchange-correlation potential
we {\it also} solve the quasiparticle equation with the frequently made 
approximation
that the quasiparticle wavefunctions are given by the Kohn-Sham 
wavefunctions. The
resulting equation for the quasiparticle energies is
\begin{equation}
\label{eq:qpeit}
  \epsilon_{i\sigma}^{qp}=\epsilon_{i\sigma}^{\rm KS}
                         +\expv{\Sigma_{\sigma\sigma}^x}
			 +\expv{\Sigma_{\sigma\sigma}^c(\epsilon_{i\sigma}^{qp})}
			 -\expv{v_{\sigma}^{xc}},
\end{equation}
where the brackets \expv{} denote matrix elements with respect to the 
Kohn-Sham wavefunction
$\phi_{i\sigma}$.

\begin{figure}
  \begin{center} 
      \epsfig{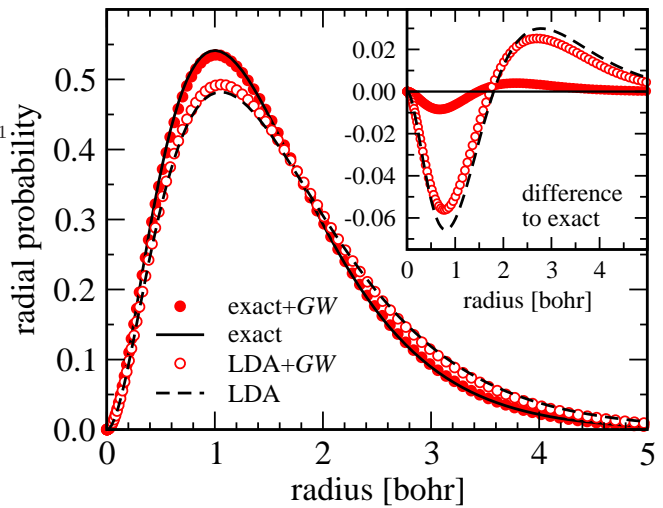}
    \caption{\label{fig:wfc} Radial probability distributions of the 
             hydrogen 1$s$ state: the quasiparticle wavefunction deviates 
             only slightly from the exact wavefunction, when the latter is 
             used as a starting point (exact+$GW$). The LDA wavefunction, 
	     on the other hand, is more delocalized as a result of 
             the inherent self-interaction.  Adding quasiparticle corrections
	     (LDA+$GW$) brings the resulting quasiparticle wavefunction 
	     slightly closer to the exact one again. The inset shows the 
	     difference to the exact wavefunction.}
  \end{center}
\end{figure}

In order to explore the role of the starting points for a $GW$ calculation, 
two
possible Kohn-Sham input Green's functions are chosen.  First, the familiar
LDA, and, second, the exact Kohn-Sham solution for the hydrogen atom which 
has the exact wavefunction of the hydrogen 1$s$ state
(\ref{eq:H_wfc}) and $v_{xc}({\bf r})=-v_H({\bf r})$.
(This exact Kohn-Sham Green's function, incidentally, differs from the
exact Green's function of the hydrogen atom, because the exact Kohn-Sham
unoccupied eigenvalues do not signify electron affinities.  The exact Green's
function cannot be constructed from any orthonormal set of one-particle
wavefunctions.)

\section{Results and discussion}
\label{section.RandD}

The calculated ionization potentials (from a full solution of the 
quasiparticle equation) are shown in Table \ref{tab:qpe}.  The 
self-interaction errors in the two
$GW$ quasiparticle energies are seen to be fairly small: 0.95 eV when the 
approximate LDA Kohn-Sham starting point is used, and the much smaller
0.21 eV when the exact Kohn-Sham starting point is used.   Clearly the LDA 
is such a physically poor representation of the correct physics in this 
extreme system (owing to the large self-interaction present in the LDA 
calculation itself, as reflected in the large error in the LDA Kohn-Sham 
eigenvalue) that it forms a very unsuitable starting point for $GW$.  
However, a physically reasonable starting point reduces the $GW$ 
self-interaction error to a small size.

\begin{table}
  \begin{ruledtabular}
    \begin{tabular}{ccccc}
      Exact & HF & LDA & LDA+$GW$ & Exact+$GW$ \\ \hline
	$-$13.61 & $-$13.61 & $-$6.36 & $-$12.66 & $-$13.40\\
    \end{tabular}
  \end{ruledtabular}
  \caption{\label{tab:qpe} Quasiparticle energies (eV) for the 1$s$ state of 
           hydrogen (the ionization potential) obtained by diagonalizing the 
	   quasiparticle Hamiltonian (\ref{qpe:full}). Two $GW$ calculations 
	   are shown, starting from the LDA and from exact Kohn-Sham, 
	   respectively. For comparison, the Hartree-Fock (HF) and LDA 
	   eigenvalues are also shown.}
\end{table}

\begin{table}
  \begin{ruledtabular}
    \begin{tabular}{lllll}
      Kohn-Sham $G_0$ &
      $\epsilon_{1s}^{GW}$ &
      $\expv{\Sigma_x}$ &
      $\expv{v_{xc}}$ &
      $\expv{\Sigma_c}$ \\ \hline
       LDA          & $-$12.93 & $-$15.38 &  $-$8.25  &  0.56 \\
       Exact        & $-$13.35 & $-$17.00 &  $-$17.00 &  0.25 \\
    \end{tabular}
  \end{ruledtabular}
  \caption{\label{tab:contrib} Quasiparticle energies (eV) for the 1$s$ 
           state of hydrogen obtained by solving Eq. (\ref{eq:qpeit}). 
	   The contributions from the exchange $\expv{\Sigma_x}$ and 
	   correlation $\expv{\Sigma_c}$ part of the self-energy are 
	   compared to that of the exchange-correlation potential
	   $\expv{v_{xc}}$ for the LDA and the exact case ($v_{xc}=-v_H$) 
	   as a starting point.  Exact value for $\epsilon_{1s}$ is $-$13.61 eV.
          }
\end{table}

Since $\expv{\Sigma_c}$ gives a non-vanishing contribution to the hydrogen 
1$s$ state, even if
the analytic solution is used as a starting point, the quasiparticle 
wavefunction will differ
from the exact one. Figure \ref{fig:wfc} shows that the $GW$ correlation 
gives rise to
a slight delocalization of the quasiparticle wavefunction in this case. This 
relaxation,
however, now makes the quasiparticle wavefunction an eigenfunction of the 
quasiparticle
Hamiltonian and reduces the deviation from the exact energy of the 1$s$ 
state to 0.2~eV, as
shown in Table \ref{tab:qpe}. In the LDA the self-interaction error is much 
more pronounced
and the wavefunction becomes significantly more delocalized. The $GW$ 
self-energy
corrects this to a small extent (as reflected in the quasiparticle 
wavefunction), but
the remaining discrepancy reiterates the unsuitability of the LDA as a 
starting point for $GW$ in this self-interaction-dominated atom.

For an analysis of the contributions to the self-energy we turn to the 
perturbative solution of the quasiparticle equation using Eq. 
\ref{eq:qpeit}, shown in Table \ref{tab:contrib}.  When the exact Kohn-Sham 
wavefunction and eigenvalues are used, as in the Hartree-Fock case
the exchange part of the self-energy is seen to cancel the self-interaction 
contribution from
the Hartree potential exactly. The correlation part, on the other hand,
is not zero, but amounts to a self-polarization of 0.25~eV. When the LDA is 
used as
starting point the influence of the LDA wavefunction on the exchange 
operator becomes
apparent and it reduces from $-$17.00~eV in the exact case to $-$15.38~eV. 
This corrects the
highly overestimated LDA eigenvalue for the 1$s$ state of $-$6.36~eV (see 
Table \ref{tab:qpe})
to $-$13.49~eV. However, in this case the contribution from the correlation 
part of the $GW$
self-energy is even larger than when starting from the exact case and 
increases the
quasiparticle energy to $-$12.93~eV.

\section{Conclusion}
\label{section.con}

We have performed spin-resolved benchmark calculations for the $GW$ 
formalism using the
analytically known solutions of the hydrogen atom as a reference, making
the self-interaction error introduced
by the correlation part of the $GW$ self-energy directly assessable.
When the exact Kohn-Sham Green's function is used as the input to $GW$, the 
self-interaction
error is small (0.21~eV, one-thirtieth the size of that in the LDA), but not 
negligible. If the LDA Kohn-Sham Green's function is used, as done in many 
$GW$ calculations for more complex systems, a larger self-interaction error 
remains, inherited from the LDA starting point.

\begin{acknowledgments}
This work was supported by the NATO Security Through Science
Programme (EAP.RIG.981521), the Slovak grant agency VEGA (project No. 
1/2020/05)
and the EU's 6th Framework Programme through the NANOQUANTA Network of 
Excellence
(NMP4-CT-2004-500198).
\end{acknowledgments}

\end{document}